\documentclass{svproc}
\usepackage[a4paper,margin=1.9cm]{geometry}
\usepackage{color}
\usepackage{bm}
\usepackage{graphicx,epsfig,psfrag,subfigure,amsopn,epstopdf,amsmath}
\usepackage{braket}
\usepackage{romannum}
\usepackage{url}

\newcommand\sg{\subfigure}
\newcommand\ig{\includegraphics}
\newcommand\be{\begin{equation}}
\newcommand\ee{\end{equation}}
\usepackage{cite}
\usepackage{xcolor}
\colorlet{linkequation}{red}
\usepackage[colorlinks]{hyperref}

\begin{document}
\mainmatter              
\title{Thermal properties of a non-Hermitian
system interacting with oscillator}
%
%

\author{Gargi Das\inst{1} \and Bhabani Prasad Mandal\inst{2}}

\authorrunning{Gargi Das et al.} 
%

%
\institute{$^{1,2}$ Department of Physics, Banaras Hindu University, Varanasi-221005, India\\
\email{$^{1}$ gargi.das@bhu.ac.in $^{2}$ bhabani@bhu.ac.in}
}

\maketitle              
\vspace{-0.4cm}
\begin{abstract}
In this work, we consider a two level $P\sigma_{z}$ pseudo-Hermitian system in contact with a thermal bath to study various thermodynamic properties. The system is realized in terms of infinitely many invariant subspaces. We find explicit solution in each subspace analytically. The quantum system undergoes a $P\sigma_{z}$ phase transition in each invariant subspace, when the non-Hermitian coupling exceeds a certain critical value ($\mu_c$). We calculate various thermodynamic quantities and observe that these quantities show divergences exactly at the exceptional points (EPs) of the theory.
\end{abstract}
\section{Introduction}
Over the past two and half decades, there has been great interest in a certain class of non-Hermitian quantum theories where the Hermiticity condition of the Hamiltonian of the system is relaxed with a physical but less constraining condition \cite{bender1998real,mostafazadeh2010pseudo}. It has been shown that a consistent quantum theory with a complete real spectrum, unitary time evolution and probabilistic interpretation for such non-Hermitian systems can be developed in a modified Hilbert space equipped with appropriate inner product \cite{mostafazadeh2003pseudo,ju2019non,tzeng2021hunting,Brody_2014}. Because of this exciting realisation, the research in non-Hermitian systems have received a huge boost over the past years \cite{mostafazadeh2010pseudo,bender2007making}. Therefore, non-Hermitian systems have found numerous applications in various branches of physics and interdisciplinary areas \cite{mostafazadeh2010pseudo,ghatak2015black,hasan2018new,hasan2018super,hasan2020role,mandal2012spectral,khare2000pt,mandal2013pt,mandal2015pt,article,PhysRevA.109.022227,pal2025parity,basu2004bound}, and some of the predictions of non-Hermitian theories are experimentally verified \cite{guo2009observation,pan2018photonic}.

Parity ($\mathcal{P}$) and Time Reversal ($\mathcal{T}$) symmetric non-Hermitian systems generally exhibit a $\mathcal{P}\mathcal{T}$-symmetry breaking transition that separates two regions: (i)$\mathcal{P}\mathcal{T}$-symmetric phase in which the entire spectrum is real and the eigenfunctions of the Hamiltonian respect $\mathcal{P}\mathcal{T}$-symmetry and (ii) $\mathcal{P}\mathcal{T}$-broken phase in which the entire spectrum (or a part of it) is in complex conjugate pairs and the eigenstates of the Hamiltonian are not the eigenstates of the $\mathcal{P}\mathcal{T}$-operator. The phase transition occurs at the EP for the particular Hamiltonian. Pseudo-Hermitian hamiltonians form another class of non-Hermitian hamiltonians satisfying: $\mathcal{H}  =S^{-1}\mathcal{H}^\dagger S,\quad S^\dagger=S$, where, S = linear Hermitian operator.

Even though the quantum theory of non-Hermitian physics is spread over all branches of physics and interdisciplinary subjects, very little has been explored in the field of information theory \cite{PhysRevA.102.862416}.  In this paper, we investigate various thermodynamic measures in a two-level system coupled to simple harmonic system in a pseudo-Hermitian manner. We show that it is possible to describe the theory in infinitely many 2-dimensional sub-Hilbert spaces. We calculate the thermodynamic quantities by calculating the partition function. We observe that these quantities diverge at the EPs. These results can be used to locate the EPs of the theory. Now, we explain the plan of the paper. In section II, we describe the model in detail. Thermodynamics of the system is described in section III. Finally results are summarised in  section IV .

\section{The Model}

   We consider a system of a spin-1/2 particle in the external magnetic field $\bf B$, coupled to an oscillator through some non-Hermitian interaction \cite{mandal2005pseudo}. For the sake of simplicity, we choose the external magnetic field in $z$-direction, ${\bf B}=B_{o}\hat{z}$ and the Hamiltonian for the system is given by:
\be
\begin{aligned}
 \mathcal{H} =  \nu \boldsymbol{\sigma}.{\bf B} +  \hbar \omega a^\dagger a + \mu(\sigma_{+}a - \sigma_{-}a^\dagger) =  \frac{\alpha}{2}\sigma_{z} + \hbar \omega a^\dagger a + \mu(\sigma_{+}a - \sigma_{-}a^\dagger) \,,
 \label{hamiltonian}
 \end{aligned}
 \ee
 
 where, \quad{}  $\alpha= 2 \mu B_{o} $ , $\mu$ is some arbitrary real parameter for non-Hermitian interaction and other parameters have usual notations. This system can also be thought of as a two-level system coupled to an oscillator where $\alpha$ is the splitting between the levels. Note that this Hamiltonian is not Hermitian as:
\begin{align}
\mathcal{H}^\dagger & =\frac{\alpha}{2}\sigma_{z} + \hbar \omega a^\dagger a - \mu(\sigma_{+}a - \sigma_{-}a^\dagger)  \neq  \mathcal{H} \,.
\end{align}

Under parity $\mathcal{P}$ and time reversal $\mathcal{T}$ transformation, we get:
\be
\begin{aligned}
\mathcal{P}\mathcal{T}\boldsymbol{\sigma}(\mathcal{P}\mathcal{T})^{-1}  =&-\boldsymbol{\sigma},\quad
\mathcal{P}\mathcal{T}{\bf B}(\mathcal{P}\mathcal{T})^{-1} = {\bf B}\,,\\
\mathcal{P}\mathcal{T}a^\dagger (\mathcal{P}\mathcal{T})^{-1}  = & a^\dagger,\quad
\mathcal{P}\mathcal{T}a(\mathcal{P}\mathcal{T})^{-1}  =  a, \quad 
\mathcal{P}\mathcal{T}\sigma_{\pm}(\mathcal{P}\mathcal{T})^{-1}  =-\sigma_{\mp}\,.
\label{P}
\end{aligned}
\ee

From Eqs.~\eqref{P}, we can see that the Hamiltonian in Eq.~\eqref{hamiltonian} is not $\mathcal{P}\mathcal{T}$ symmetric: $\mathcal{P}\mathcal{T} \mathcal{H} (\mathcal{P}\mathcal{T})^{-1} \neq \mathcal{H}$

However, this Hamiltonian is pseudo-hermitian with respect to $\sigma_{z}$  as well as $\mathcal{P}$ as follows:
\be
\begin{aligned}
\sigma_{z} \mathcal{H} \sigma_{z}^{-1} & = \frac{\alpha}{2}\sigma_{z} + \hbar \omega a^\dagger a + \mu(\sigma_{z}\sigma_{+}\sigma_{z}a -\sigma_{z}\sigma_{-}\sigma_{z}a^\dagger) = \mathcal{H}^\dagger \,,\\
\mathcal{P H } \mathcal{P}^{-1}& = \frac{\alpha}{2}P\sigma_{z}P^{-1} + \hbar \omega P a^\dagger a P^{-1} + \mu(P\sigma_{+}a P^{-1} - P\sigma_{-}a^\dagger P^{-1}) 
 = \mathcal{H}^\dagger \,.
\end{aligned}
\ee

So, the Hamiltonian is invariant under the symmetry
generated by the combined operator, $P\sigma_{z}$: $[\mathcal{H},P\sigma_{z}]=0$

To find the energy eigenvalues and the corresponding eigenvectors of the system
described by the Hamiltonian in Eq.~\eqref{hamiltonian}, we adopt the notation for the state as, $|n,\frac{1}{2}m_{s}\rangle$ where: $a^\dagger a|n\rangle  = n|n\rangle,\ \sigma_{z}|\frac{1}{2}m_{s}\rangle  =m_s|\frac{1}{2}m_{s}\rangle$.

It is readily seen that $|0,-\frac{1}{2}\rangle$ is a ground state of the Hamiltonian with eigenvalue $-\frac{\alpha}{2}$ and it is non-degenerate, i.e., $\mathcal{H}|0,-\frac{1}{2}\rangle=-\frac{\alpha}{2}|0,-\frac{1}{2}\rangle$.

Note that $|n,\pm\frac{1}{2}\rangle$ are not the eigenstates of the system.
However, we observe that the states $|n,\frac{1}{2}\rangle$ and $|n+1,-\frac{1}{2}\rangle$ form an invariant subspace,
wherein the Hamiltonian matrix is given by:
\be\label{H}
\mathcal{H}_{n+1}=
\begin{pmatrix}
\frac{\alpha}{2}+n\hbar\omega & \mu\sqrt{n+1} \\
-\mu\sqrt{n+1} & -\frac{\alpha}{2}+(n+1)\hbar\omega
\end{pmatrix}\,.
\ee

Next, we obtain the Hermitian conjugate of the Hamiltonian matrix in Eq.~\eqref{H}:
\be\label{Hd}
\mathcal{H}_{n+1}^\dagger=
\begin{pmatrix}
\frac{\alpha}{2}+n\hbar\omega & -\mu\sqrt{n+1} \\
\mu\sqrt{n+1} & -\frac{\alpha}{2}+(n+1)\hbar\omega
\end{pmatrix}\,.
\ee

We calculate the eigenvalues of the $\mathcal{H}_{n+1}$ matrix in Eq.~\eqref{H} $(R^{\Romannum{1},\Romannum{2}}_{n+1})$ and $\mathcal{H}_{n+1}^\dagger$ in Eq.~\eqref{Hd}  $(L^{\Romannum{1},\Romannum{2}}_{n+1})$ to be the same as follows:
\be
\begin{aligned}\label{energy_eigenvalues}
R^{\Romannum{1},\Romannum{2}}_{n+1} = & L^{\Romannum{1},\Romannum{2}}_{n+1}
=  \frac{1}{2}[(2n+1)\hbar\omega\pm\sqrt{(\hbar\omega-\alpha)^2-4\mu^2(n+1)}]\,.
\end{aligned}
\ee

The region where eigenvalues are real is called the unbroken region, whereas the region where eigenvalues are complex is called the broken region. Therefore, from Eq.~\eqref{energy_eigenvalues}, we derive the conditions for both unbroken and broken regions as follows:
\be\label{condition}
\begin{aligned}
&|\hbar\omega-\alpha|\geq 2\mu\sqrt{n+1} \qquad  for \ unbroken \ region \,,\\
&|\hbar\omega-\alpha|< 2\mu\sqrt{n+1} \qquad  for \ broken \ region \,.
\end{aligned}
\ee

Therefore, the eigenvalues coalesce at  $\mu=\mu_c$ for our system, given by (using Eq.~\eqref{energy_eigenvalues}), for the ${(n+1)}^{th}$ subspace:
\be\label{critical}
\mu_{c_{n+1}}= \frac{|\hbar\omega-\alpha|}{2\sqrt{n+1}}\,.
\ee

\subsection{$\eta$ for Unbroken region}
For unbroken region, we follow the condition in Eq.~\eqref{condition} and proceed as follows.

We calculate the normalised eigenvectors for the Hamiltonian $\mathcal{H}_{n+1}$  $(|R^{\Romannum{1},\Romannum{2}}_{n+1}\rangle)$ and those for its Hermitian conjugate matrix, $\mathcal{H}_{n+1}^\dagger$ $(|L^{\Romannum{1},\Romannum{2}}_{n+1}\rangle)$ in Eqs.~\eqref{H} and ~\eqref{Hd} using the bi-orthonormality condition. Next, we calculate the $\eta$ matrix in the unbroken region for the ${(n+1)}^{th}$ subspace as follows and substituting $\hbar=\omega=1, \alpha=5$):
\be 
\begin{aligned}\label{eta}
\eta_{n+1} =  \sum_{i} |L^i_{n+1}\rangle\langle L^i_{n+1}|
= 
\begin{pmatrix}
\frac{2}{\sqrt{4-(1+n) \mu^2}} & \frac{\sqrt{(1+n)} \ \mu}{\sqrt{4-(1+n) \mu^2}}  \\
\frac{\sqrt{(1+n)} \ \mu}{\sqrt{4-(1+n) \mu^2}}  & \frac{2}{\sqrt{4-(1+n) \mu^2}} 
\end{pmatrix}\,.
\end{aligned}
\ee

\subsection{$\eta$ for Broken region}
The region where eigenvalues are complex is called the broken region.
For broken region, we follow the condition in Eq.~\eqref{condition} and proceed as follows.

We find the normalised eigenvectors for the Hamiltonian $\mathcal{H}_{n+1}$ $(|\tilde{R}^{\Romannum{1},\Romannum{2}}_{n+1}\rangle)$ and those for its Hermitian conjugate matrix $\mathcal{H}_{n+1}^\dagger$ $(|\tilde{L}^{\Romannum{1},\Romannum{2}}_{n+1}\rangle)$ in Eqs.~\eqref{H} and ~\eqref{Hd} using the bi-orthonormality condition. We calculate the $\eta$ matrix in the broken region for the ${(n+1)}^{th}$ subspace as follows  and substituting $\hbar=\omega=1, \alpha=5$):
\be 
\begin{aligned}\label{etad}
\tilde{\eta}_{n+1} =  \sum_{i} |\tilde{L}^i_{n+1}\rangle\langle \tilde{L}^i_{n+1}|
= 
\begin{pmatrix}
\frac{\sqrt{(1+n)}\ \mu}{\sqrt{(1+n) \mu^2-4}} & \frac{2}{\sqrt{(1+n) \mu^2-4}}  \\
\frac{2}{\sqrt{(1+n) \mu^2-4}}  &\frac{\sqrt{(1+n)}\ \mu}{\sqrt{(1+n) \mu^2-4}} 
\end{pmatrix}\,.
\end{aligned}
\ee

\section{Thermodynamics for the system}

To calculate various thermodynamic properties, we consider this system in contact of a thermal bath at equilibrium temperature, T. The partition function Z of a system is given as,

\be
Z=tr\left(e^{-H/k_{B}T}\right)=tr\left(e^{-H/\tau}\right)\,.
\ee

Since, our system is non-Hermitian, so the partition function for our model gets modified as follows \cite{cao2023statistical}:

\be
Z=tr\left(e^{-H/\tau} \eta\right)\,.
\ee

For our system:

In the unbroken region, the partition function for the ${(n+1)}^{th}$ subspace is given by (using Eqs.~\eqref{H} and ~\eqref{eta} ):
\be\label{pf_n}
Z_{n+1}= tr\left(e^{-H_{n+1}/\tau} \eta_{n+1}\right)\,.
\ee

Whereas in the broken region, the partition function for the ${(n+1)}^{th}$ subspace is given by (using Eqs.~\eqref{H} and ~\eqref{etad} ):
\be\label{pfb_n}
\tilde{Z}_{n+1}= tr\left(e^{-H_{n+1}/\tau} \tilde{\eta}_{n+1}\right)\,.
\ee

Next, we calculate various thermodynamic quantities such as the Free Energy $F$, Entropy $S$, and
Specific Heat $C_{v}$, for the system using partition function as follows:

\begin{align}
    F  =-\tau\ln Z, \qquad
    \frac{S}{k_B}  =\ln Z +\tau\frac{\partial\ln Z}{\partial\tau}, \qquad
    \frac{C_v}{k_B} = 2\tau\frac{\partial\ln Z}{\partial\tau}+\tau^2 \frac{\partial^2\ln Z}{\partial\tau^2}\label{S}\,.
\end{align}

\begin{figure}[htb]
\vspace{-0.4cm}
\begin{center}
\sg[~$n=0$]{\ig[height=2.9cm]{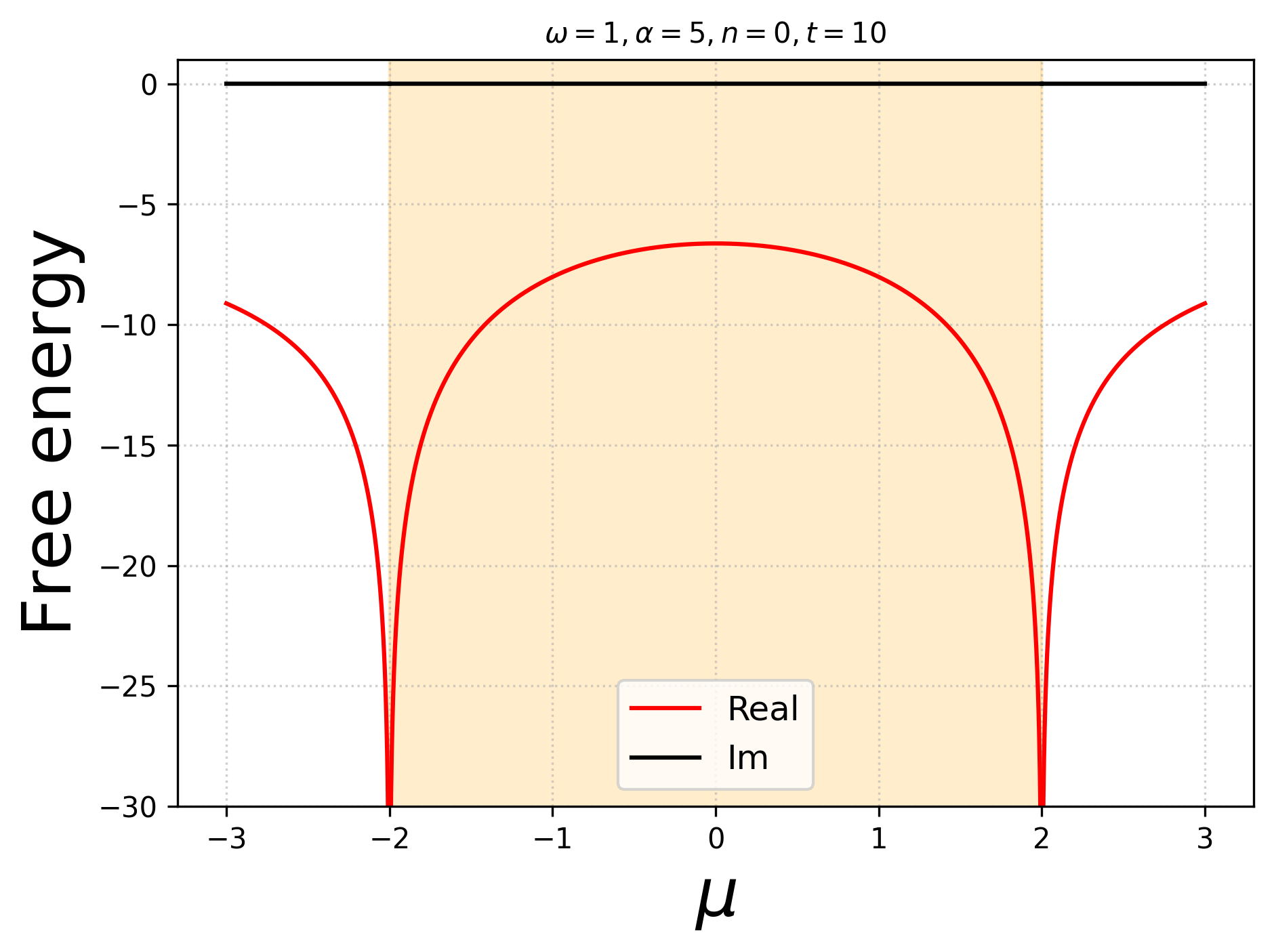}\label{fig:2a}}
\sg[~$n=1$]{\ig[height=2.9cm]{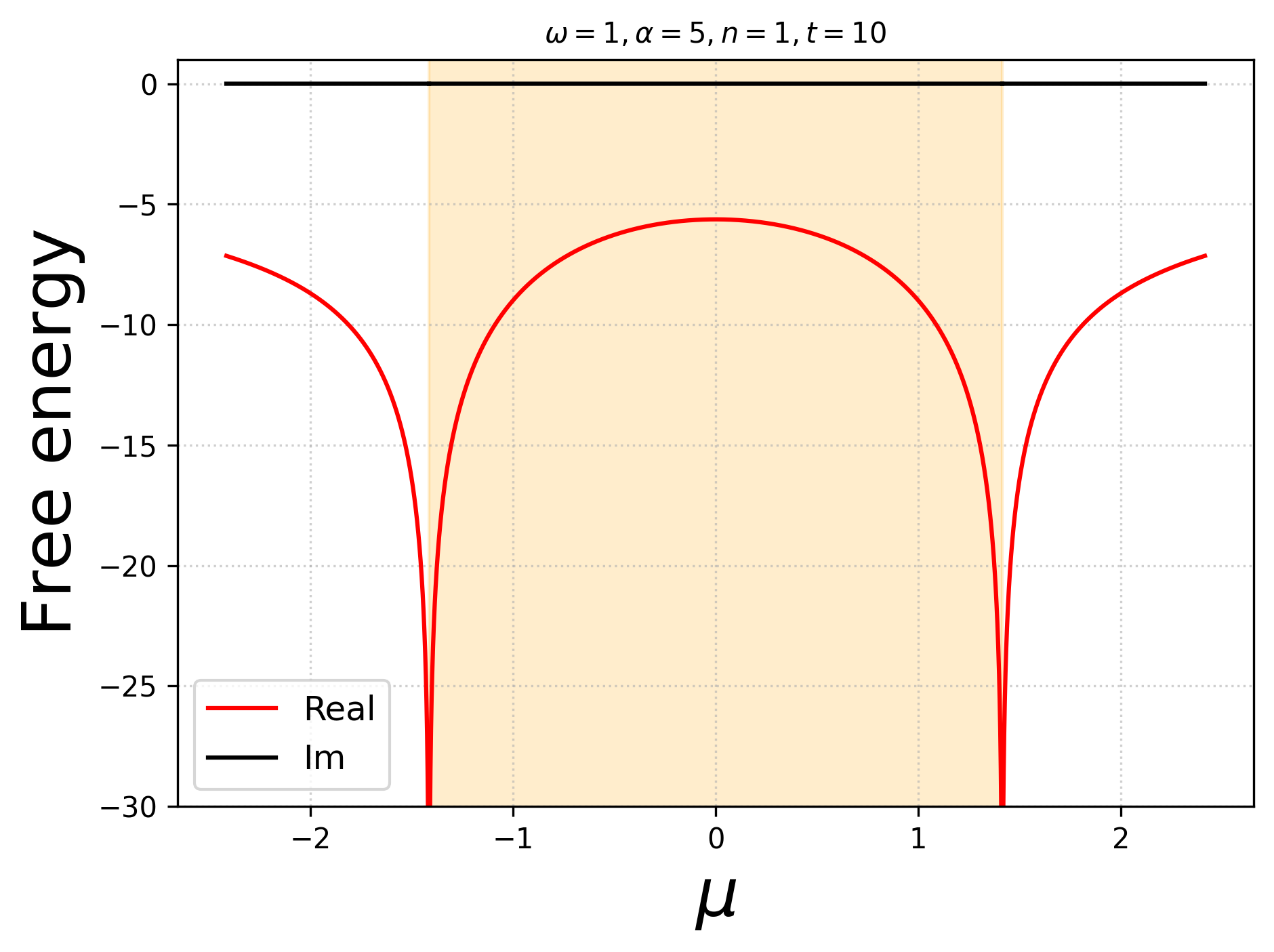}\label{fig:2b}}
\sg[~$n=2$]{\ig[height=2.9cm]{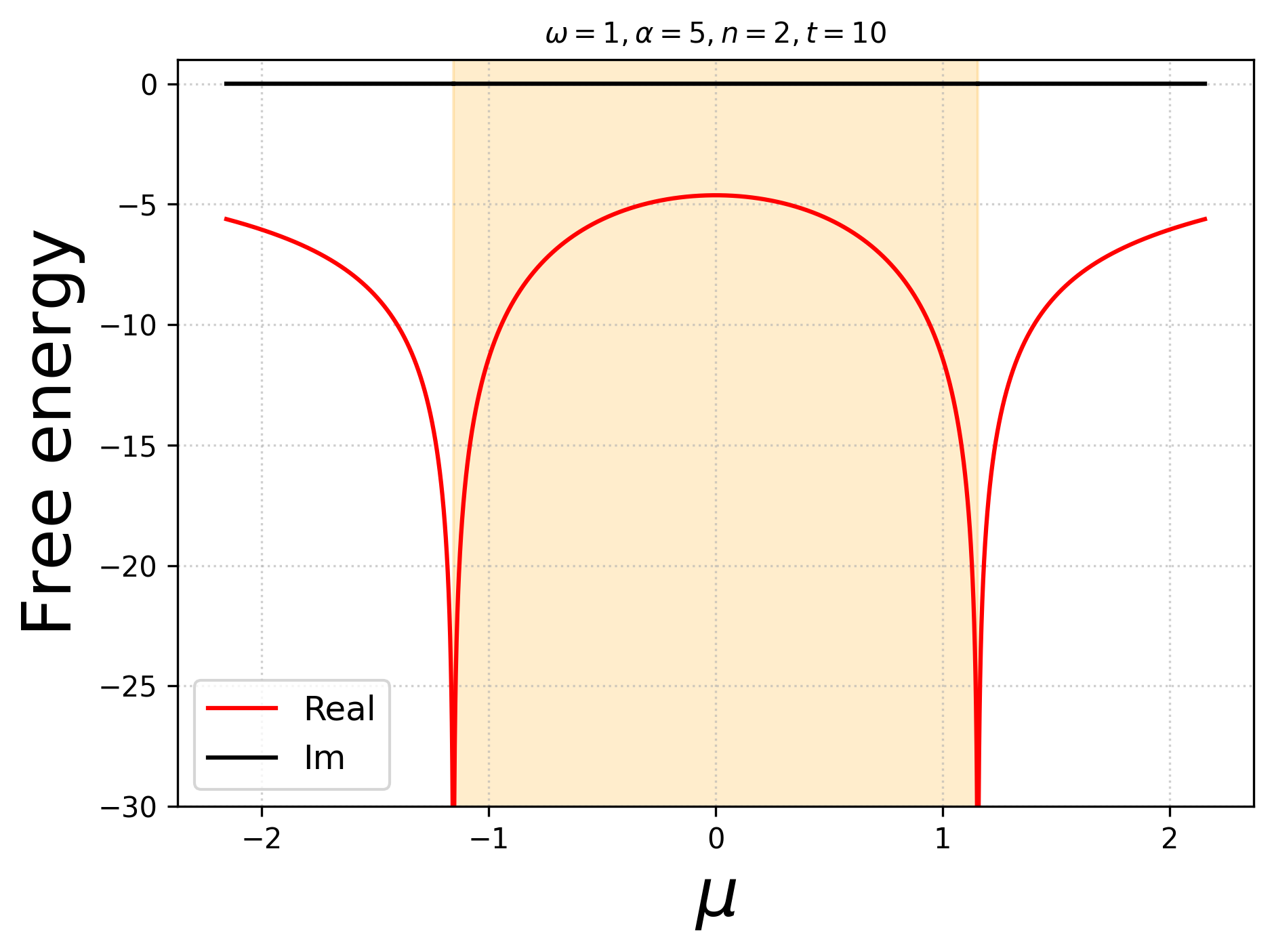}\label{fig:2c}}
\sg[~$n=5$]{\ig[height=2.9cm]{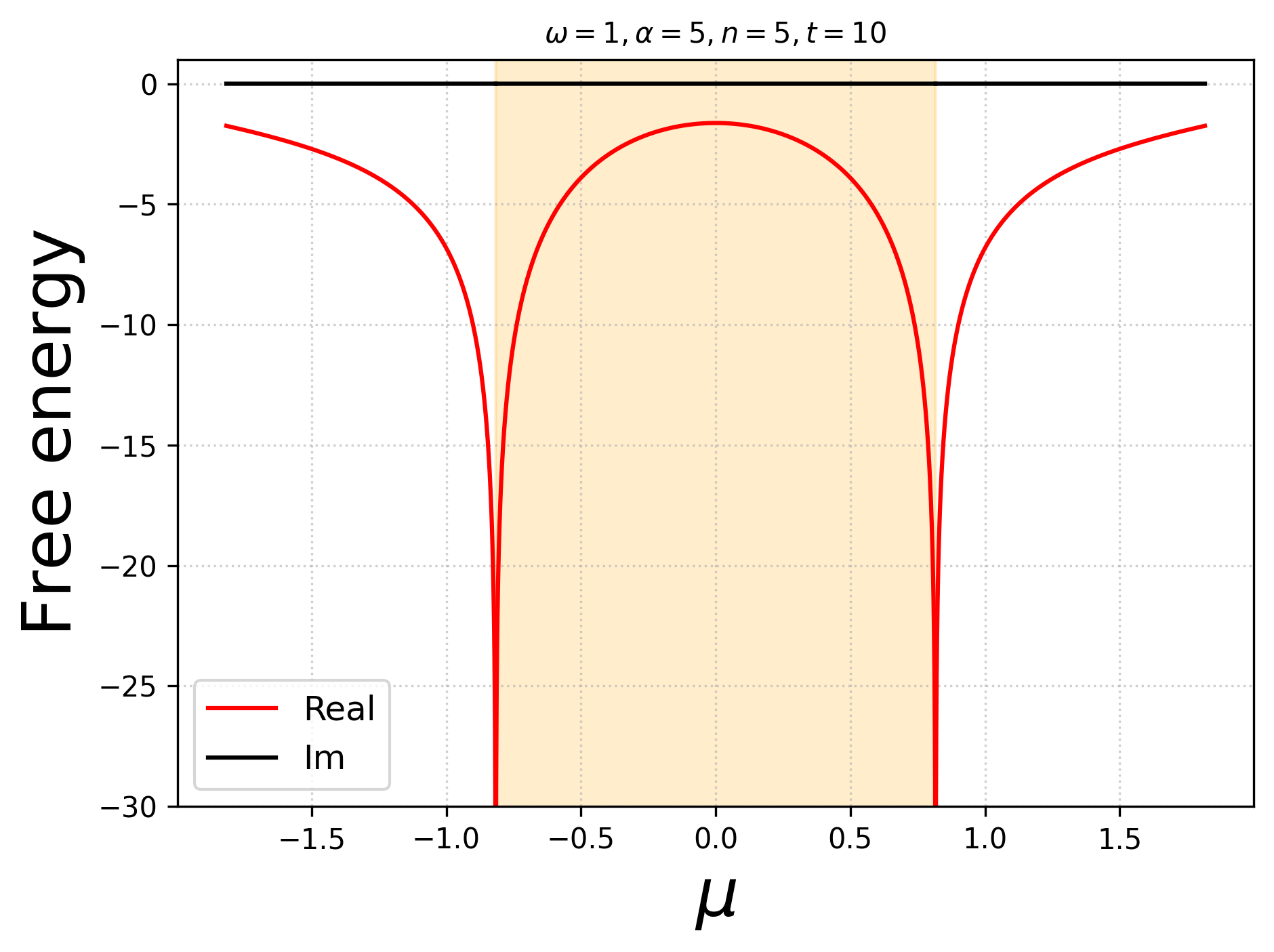}\label{fig:2d}}
\end{center}
\vspace{-0.4cm}
\caption{{Variation of the Free energy as a function of the real parameter, $\mu$ for different values of n  ($\hbar=\omega=1, \alpha=5$). The shaded region denotes the unbroken region, whereas the unshaded region denotes the broken region}} 
\label{fig:1} 
\end{figure}

\begin{figure}[htb]
\begin{center}
\sg[~$n=0$]{\ig[height=2.9cm]{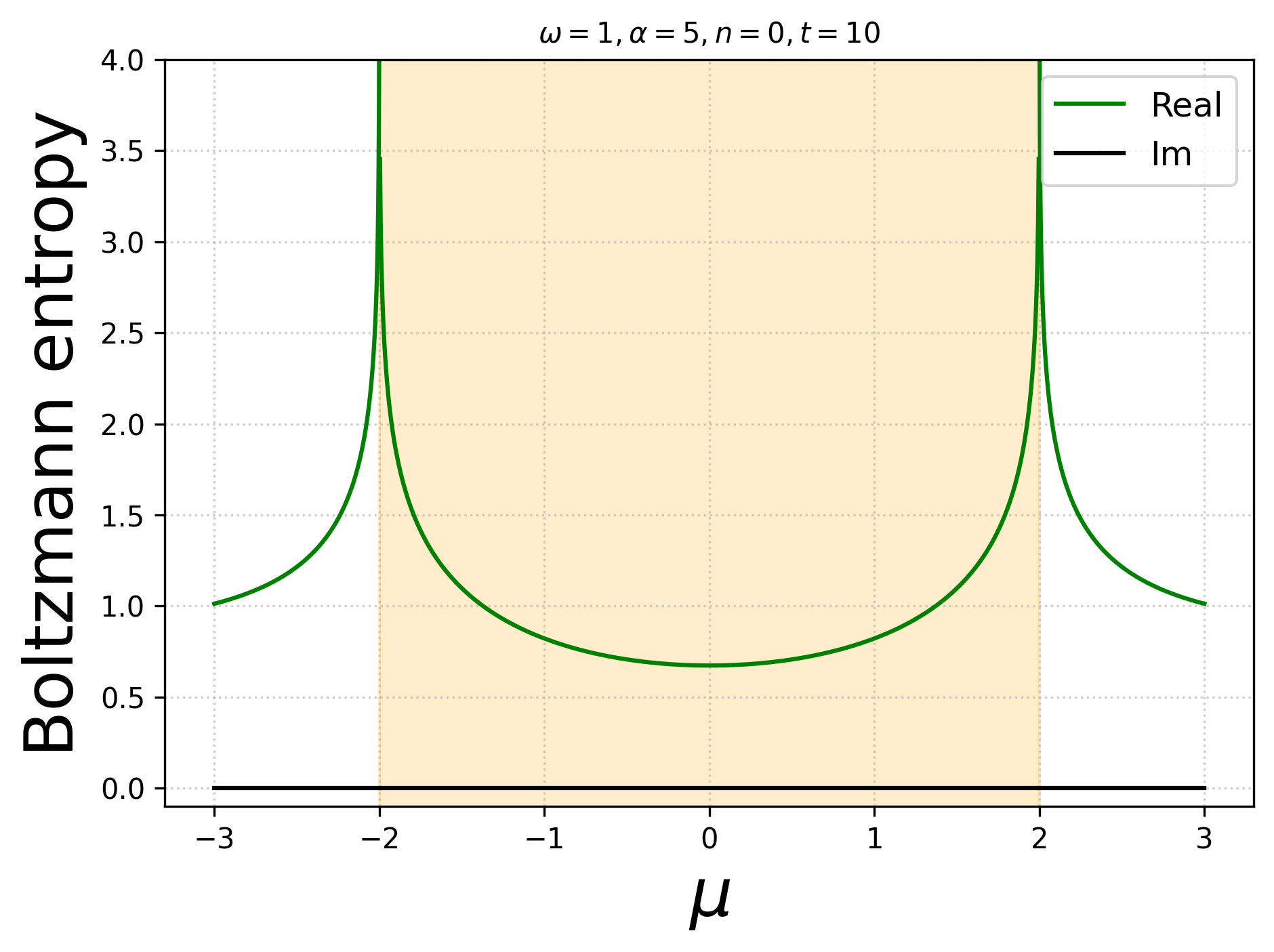}\label{fig:2a}}
\sg[~$n=1$]{\ig[height=2.9cm]{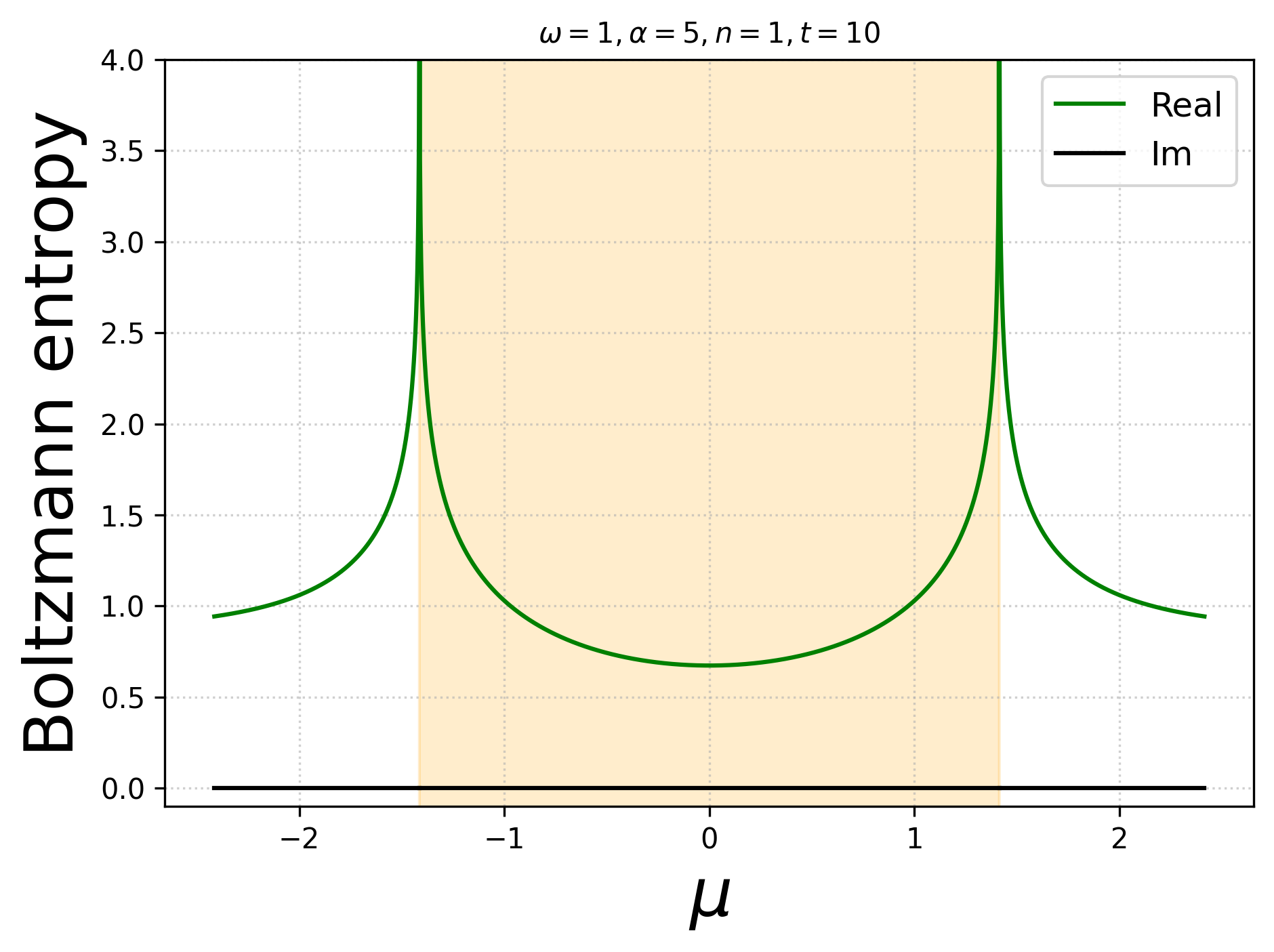}\label{fig:2b}}
\sg[~$n=2$]{\ig[height=2.9cm]{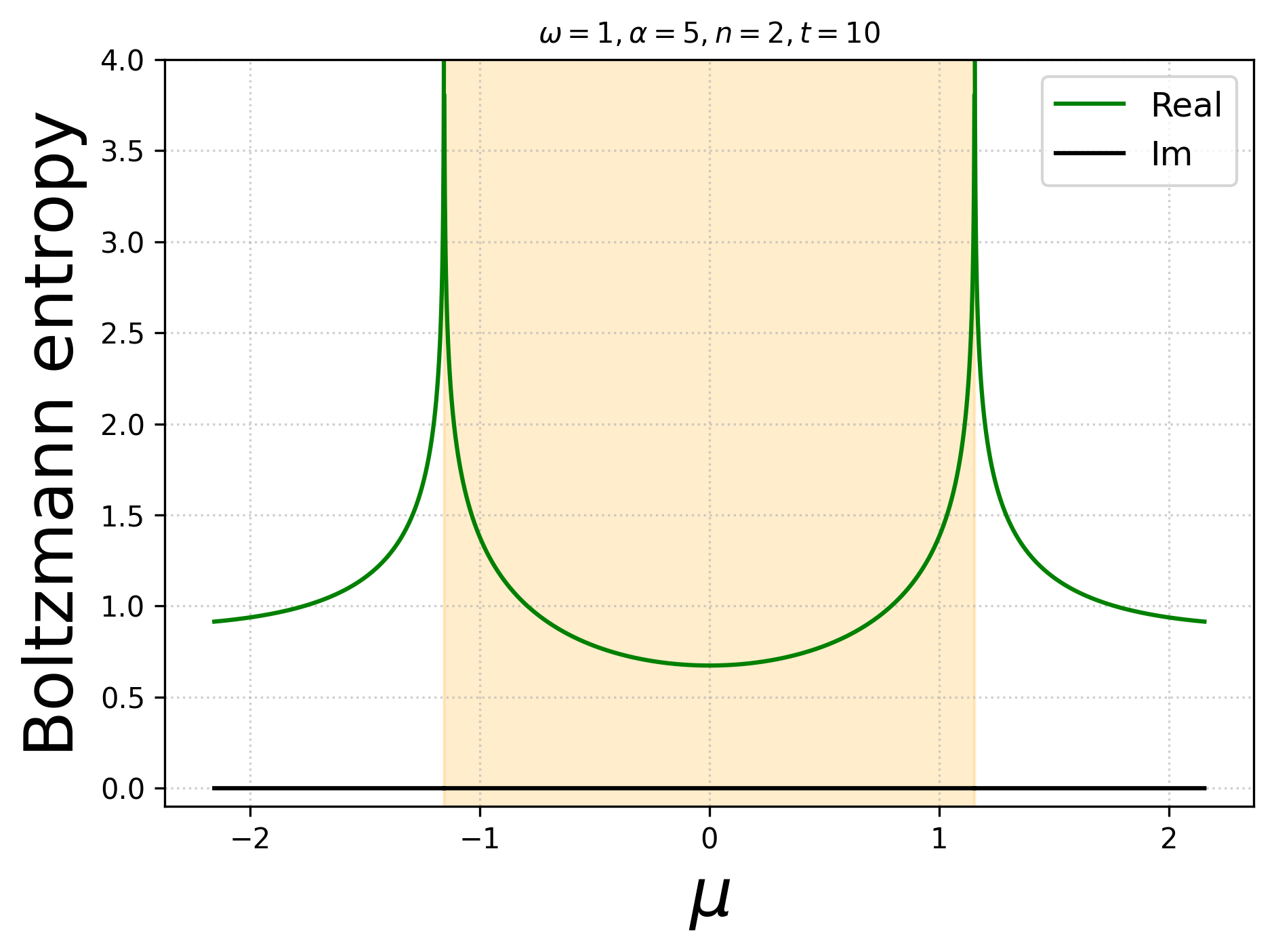}\label{fig:2c}}
\sg[~$n=5$]{\ig[height=2.9cm]{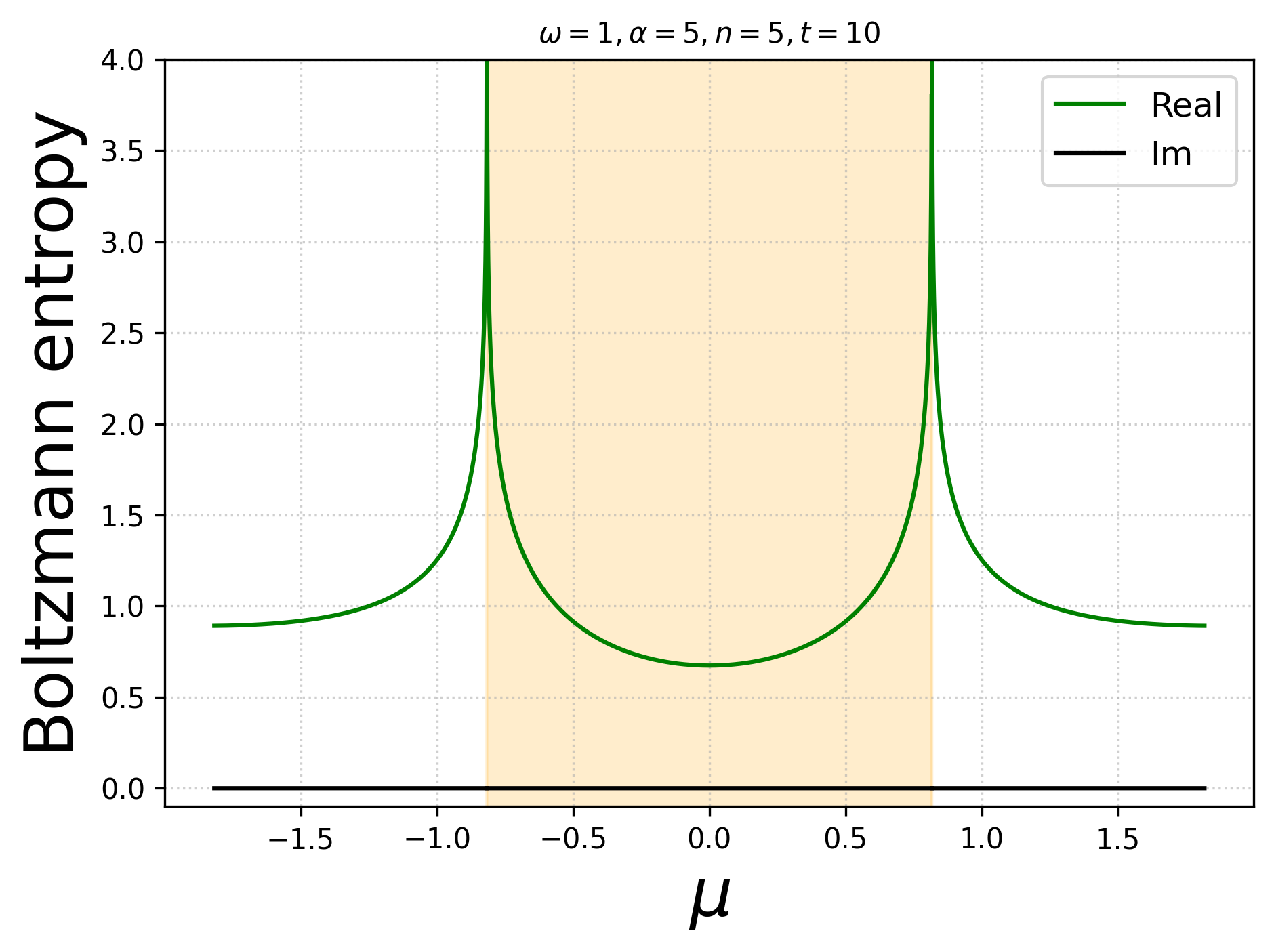}\label{fig:2d}}
\end{center}
\vspace{-0.4cm}
\caption{{Variation of the Boltzmann entropy as a function of the real parameter, $\mu$ for different values of n  ($\hbar=\omega=1, \alpha=5$). The shaded region denotes the unbroken region, whereas the unshaded region denotes the broken region}} 
\label{fig:2} 
\vspace{-0.9cm}
\end{figure}

\begin{figure}[htb]
\begin{center}
\sg[~$n=0$]{\ig[height=2.9cm]{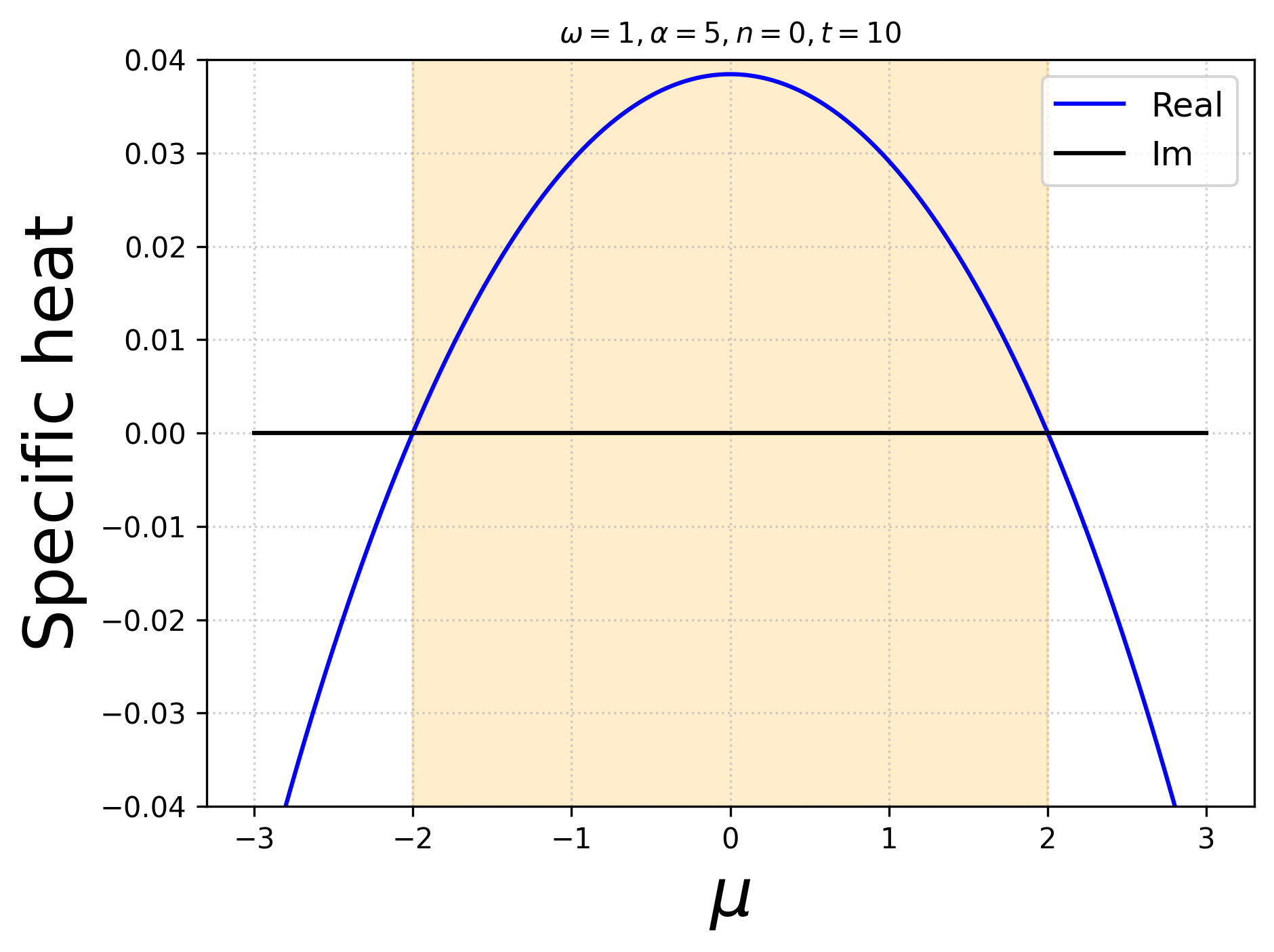}\label{fig:2a}}
\sg[~$n=1$]{\ig[height=2.9cm]{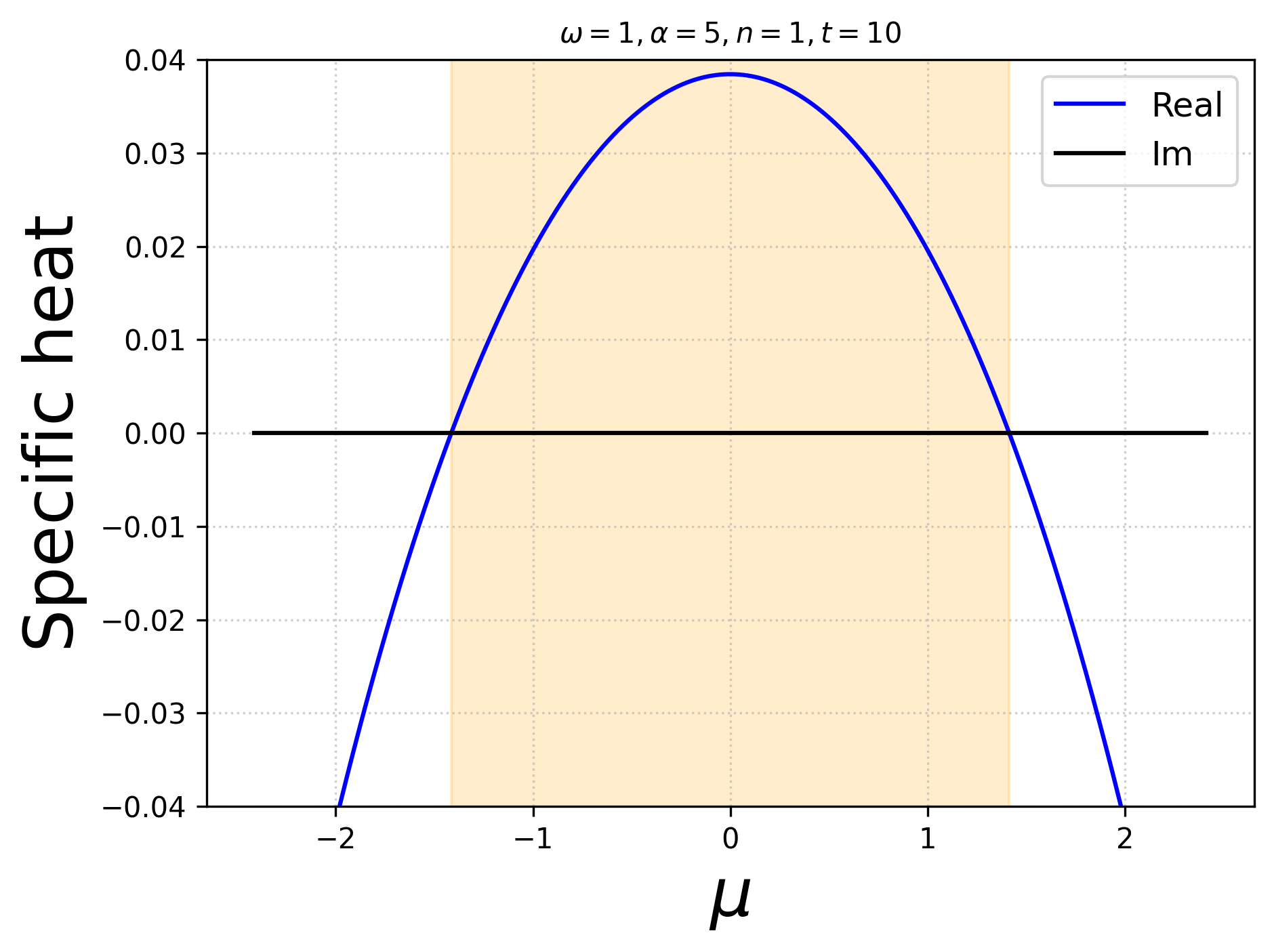}\label{fig:2b}}
\sg[~$n=2$]{\ig[height=2.9cm]{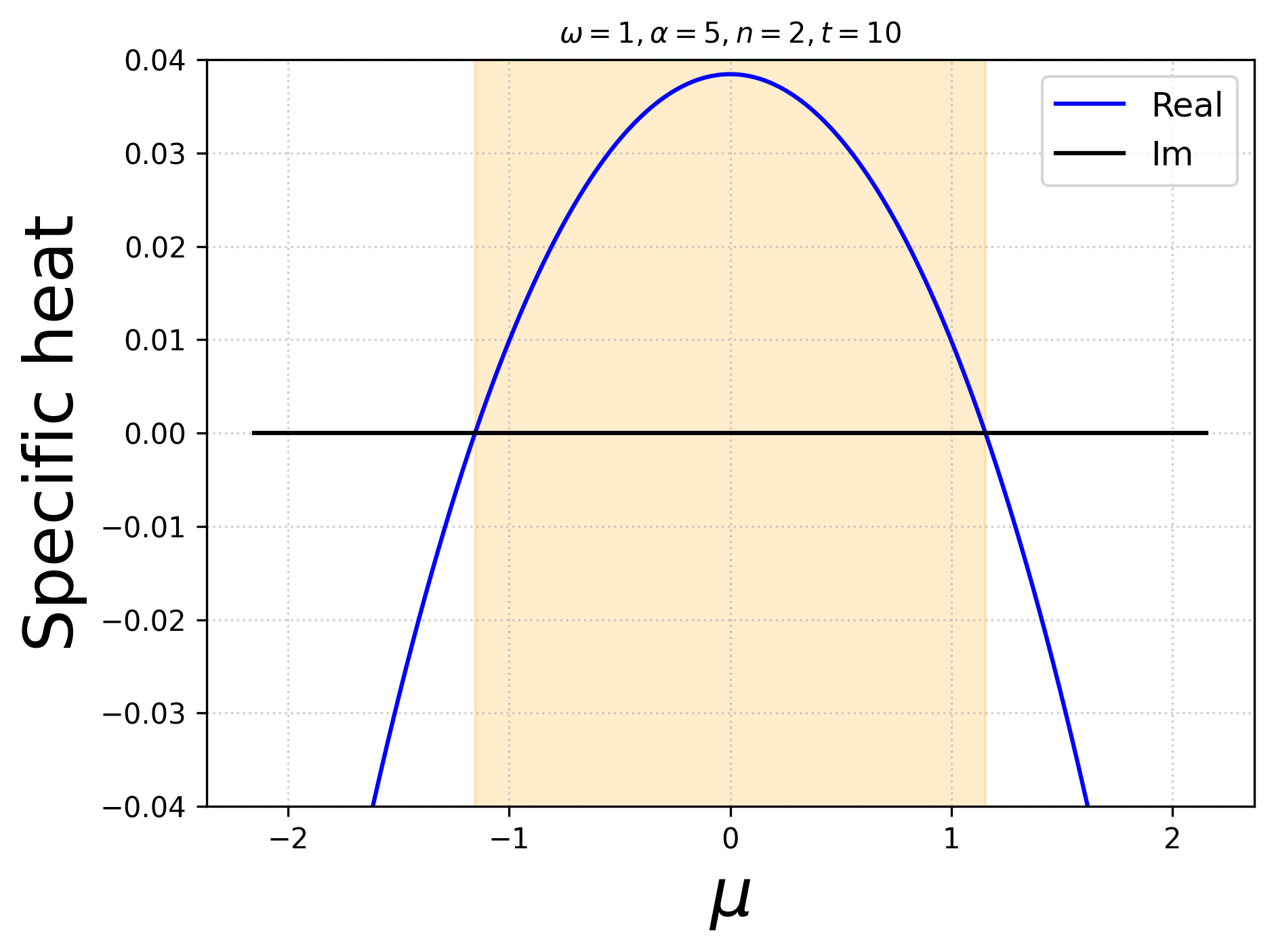}\label{fig:2c}}
\sg[~$n=5$]{\ig[height=2.9cm]{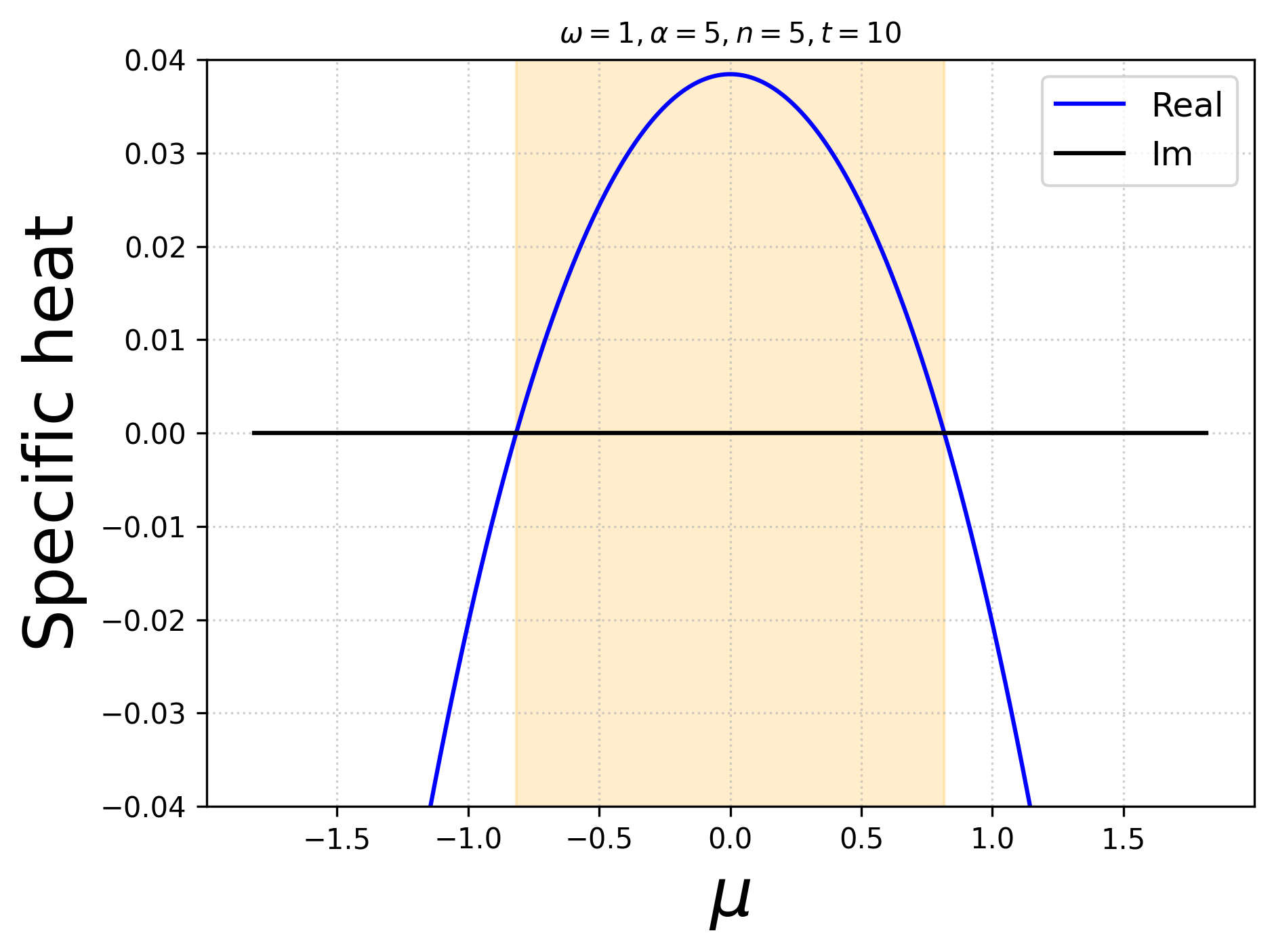}\label{fig:2d}}
\end{center}
\caption{{Variation of the Specific heat as a function of the real parameter, $\mu$ for different values of n  ($\hbar=\omega=1, \alpha=5$). The shaded region denotes the unbroken region, whereas the unshaded region denotes the broken region}} 
\label{fig:3} 
\end{figure}

Using Eqs.~\eqref{pf_n}-\eqref{S}, we calculate the thermodynamic quantities for both unbroken and broken regions. The variation of the thermodynamic quantities with respect to $\mu$ are shown in Figs. (\ref{fig:1})-(\ref{fig:3}) for different subspaces.

\vspace{-1.2cm}

\section{Discussion and Results}
 We consider a two level $P\sigma_{z}$ pseudo-Hermitian system in contact with a thermal bath to study various thermodynamic properties. We observe that the free energy and Boltzmann entropy are completely real in both unbroken and broken regions, and they diverge at the EPs for each subspace as shown in Fig. (\ref{fig:1}) and Fig. (\ref{fig:2}) respectively. Also, we observe that  the Specific heat remains positive in the unbroken region while they take negative (unphysical) values in the broken region and become zero at the EPs for each subspace as shown in Fig. (\ref{fig:3}). From the above discussion, we conclude that all the thermodynamic quantities are real in all regions. It is also visible that these quantities show specific behaviour at the EPs  for each subspace. Further, we can investigate various measures for quantum information for this model.

\section{Acknowledgment}
BPM acknowledges the incentive research grant for faculty under the IoE Scheme (IoE/Incentive/2021-22/32253) of Banaras Hindu University. G.D. acknowledges UGC, New Delhi, India for JRF fellowship.

%
%

\end{document}